\documentclass[12pt]{article}

\usepackage{setspace}
\usepackage{float}
\usepackage{amssymb}
\usepackage{amsthm}
\usepackage{amsmath}
\usepackage{latexsym}
\usepackage{epsfig}
\usepackage{graphicx}
\usepackage{wasysym}
\usepackage{threeparttable}
\usepackage{natbib}
\usepackage{color}
\usepackage{epstopdf}
\usepackage{caption}
\usepackage{subcaption}

\usepackage[breaklinks]{hyperref}

\def\boxit#1{\vbox{\hrule\hbox{\vrule\kern6pt
          \vbox{\kern6pt#1\kern6pt}\kern6pt\vrule}\hrule}}

\def\bse{\begin{eqnarray*}}
\def\ese{\end{eqnarray*}}
\def\be{\begin{eqnarray}}
\def\ee{\end{eqnarray}}
\def\bq{\begin{equation}}
\def\eq{\end{equation}}
\def\bse{\begin{eqnarray*}}
\def\ese{\end{eqnarray*}}

\usepackage{mathptmx}      
\usepackage{bm}

 {\begin{list}{}%
         {\setlength{\leftmargin}{#1}}%
         \item[]%
 }
 {\end{list}}

\newcommand{\bdm}[1]{\mathbf{#1}}

\newcommand{\ser}[2]{${#1}_1,{#1}_2,\cdots,{#1}_{#2}$} 

\usepackage{amsmath}

\newtheorem{theorem}{Theorem}
\newtheorem{corollary}{Corollary}[theorem]
\usepackage{subcaption}

\usepackage[export]{adjustbox}

\begin{document}
\thispagestyle{empty} \baselineskip=28pt

\begin{center}
	{\LARGE{\bf Constrained Bayesian Hierarchical Models for Gaussian Data: A Model Selection Criterion Approach
	}}
\end{center}

\baselineskip=12pt

\vskip 2mm
\begin{center}
Qingying Zong\footnote{(\baselineskip=10pt to whom correspondence should be addressed) Department of Statistics, Florida State University, 117 N. Woodward Ave, Tallahassee, Fl 32306, qingying.zong@stat.fsu.edu}, and Jonathan R. Bradley\footnote{Department of Statistics, Florida State University, 117 N. Woodward Ave, Tallahassee, Fl 32306, qingying.zong@my.fsu.edu}
\end{center}
%
%
%
%
\vskip 4mm

\begin{center}
\large{{\bf Abstract}}
\end{center}
\noindent
Consider the setting where there are $B$ $(\ge 1)$ candidate statistical
models, and one is interested in model selection. Two common approaches to solve this problem are to select a single model or to combine the candidate models through model averaging. Instead, we select a subset of the combined parameter space
associated with the models. Specifically, a model averaging perspective is used to increase the parameter space, and a model selection criterion is used to select a subset of this expanded parameter space. We account for the variability of the criterion by adapting \citet{yekutieli2012adjusted}'s method
to Bayesian model averaging (BMA). \citet{yekutieli2012adjusted}'s method treats model selection as a truncation problem. We truncate the joint support of the data and the parameter space to only include
small values of the covariance penalized error (CPE) criterion. The CPE is a general expression that contains several information criteria as special cases. Simulation results show that as long as the truncated set does not have near zero probability, we tend to obtain lower mean
squared error than BMA. Additional theoretical results are provided that provide the foundation for these observations. We apply our approach to a dataset consisting of American Community Survey (ACS) period estimates to illustrate that this perspective can lead to improvements of a single model.
\baselineskip=12pt

%
%
%

\baselineskip=12pt
\par\vfill\noindent
{\bf Keywords:} {Bayesian hierarchical model; Markov chain Monte Carlo; Posterior predictive p-value; Information theory; Gaussian Processes.
\par\medskip\noindent
\clearpage\pagebreak\newpage \pagenumbering{arabic}
\baselineskip=24pt

\section{Introduction}\label{sec:intro}

The goal of statistical  model selection is often to either select one model (see for example, \citealt{akaike1973maximum}) or combine the candidate models \citep{hoeting1999bayesian}. In this article, we combine both types of perspectives. In particular, a model averaging perspective is used to increase the parameter space, and a selection criterion is used to select a subset of this expanded parameter space. The selected set is chosen so that the values in the set have high predictive performance. See Figure \ref{circle1} for an illustration of this new perspective. Use of selection  criteria often results in the selection of a single model among all the competing models (e.g., the green shaded region in left panel of Figure \ref{circle1}, or $M_1$). Our approach is similar, but the selected subset is not restricted to a single model (e.g., the black triangle region is an abstract representation of our selected model in left panel of Figure \ref{circle1}).

\begin{figure}[hbt!]
	\centering
	\begin{subfigure}{.6\textwidth}
		\centering
		\includegraphics[width=1.3\linewidth,center]{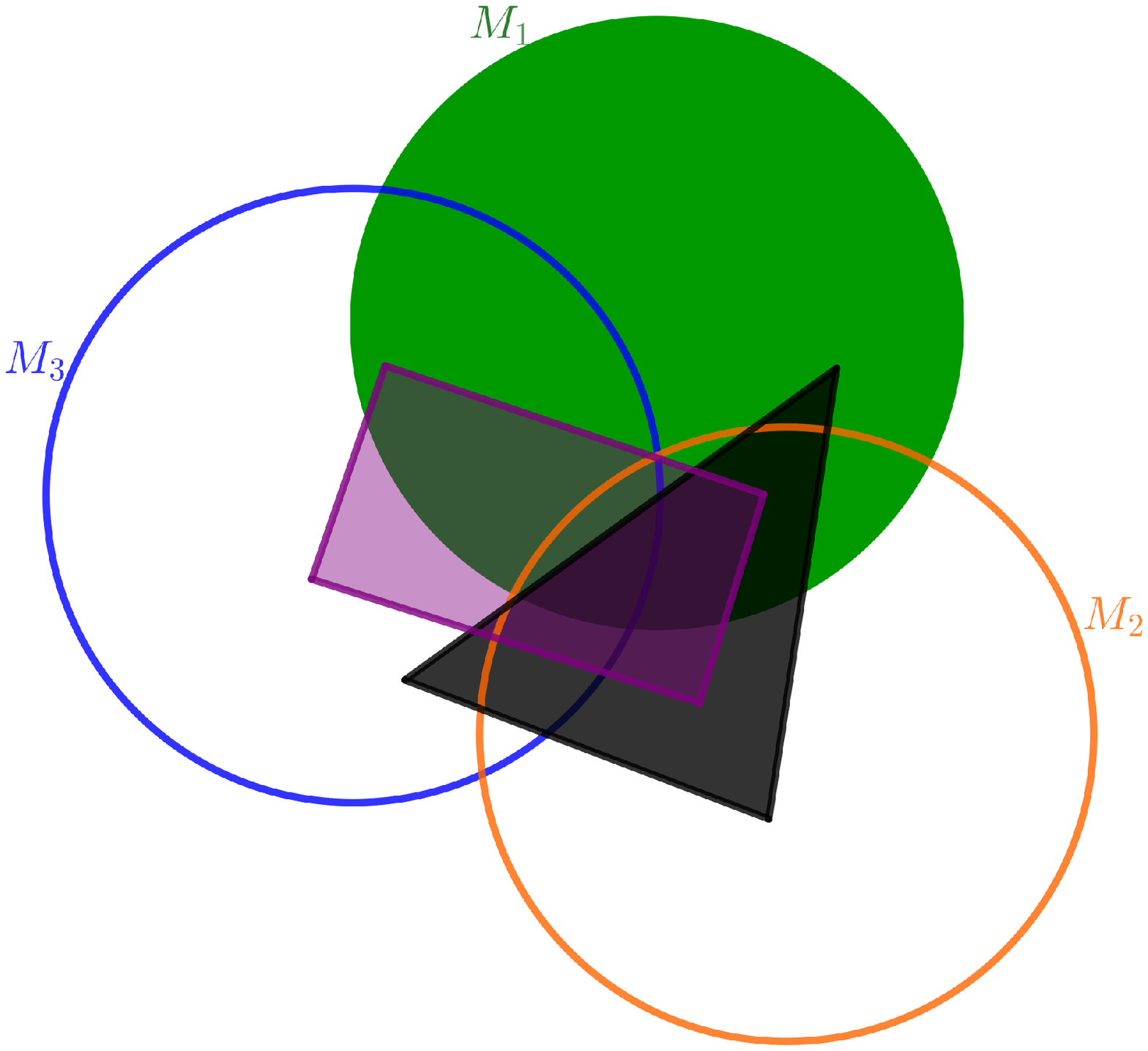}
	\end{subfigure}%
	\begin{subfigure}{.6\textwidth}
		\includegraphics[width=1.3\linewidth,right]{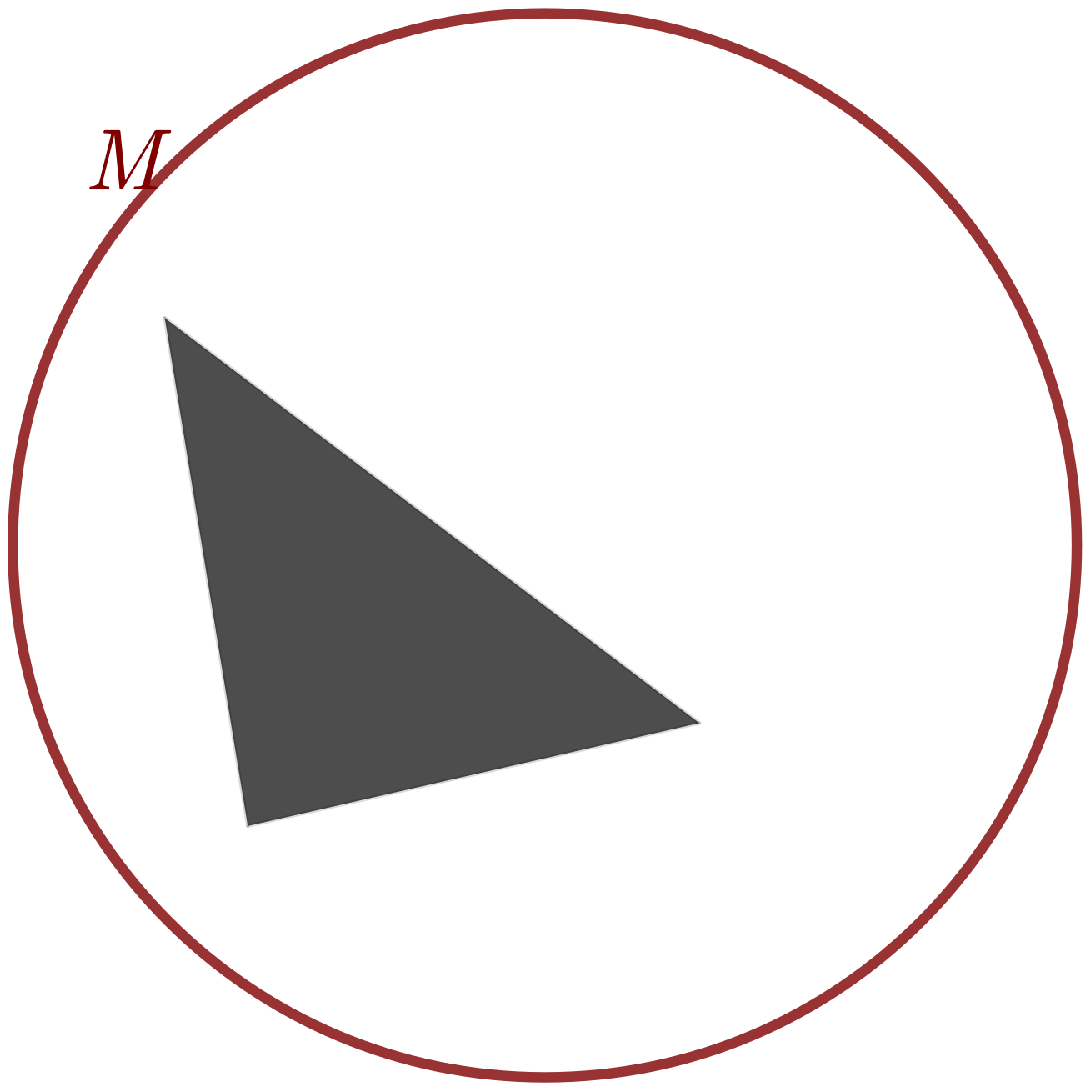}
	\end{subfigure}
	\caption{$M_1$, $M_2$, and, $M_3$ are abstract representations of the parameter space of three candidate statistical models. Their union is the "expanded parameter space," the black shaded triangle region represents values associated with high predictive performance, and the purple shaded rectangle provides an abstract representation of the posterior distribution based on the sparsity inducing priors.}
	\label{circle1}
\end{figure}

This combined perspective is similar to the use of sparsity inducing priors. (See the purple shaded rectangle in the left panel Figure \ref{circle1}, which provides an abstract representation of the use of sparsity inducing priors.) The difference with our approach is that we are selecting a region we believe to have high predictive performance through the use of selection criteria, where sparsity inducing priors use Bernoulli \citep[e.g., see][for the spike and slab prior]{ishwaran2005spike} or ``near Bernoulli'' priors \citep[e.g., see][for the horsehoe prior]{carvalho2009handling} to effectively select a subset of an expanded parameter space.

Combining these two perspectives (i.e., BMA and selection criteria) in the proposed way removes some of the inferential issues with the individual perspectives. For example, BMA accounts for the variability in selecting a model, but enlarges the parameter space (e.g., left panel of Figure \ref{circle1}). This perspective incorporates all potential models, but does not reduce the parameter space, and instead BMA increases the parameter space. This approach, while principled, does not result in parsimony (i.e., a smaller model). As a consequence, our summaries are relegated to an average of this larger parameter space, which includes poor performing models. For example, \citet{wasserman2000bayesian} estimates a quantity under each candidate model and then averaging the estimates with respect to how probable each model is. The aforementioned BMA \citep{hoeting1999bayesian} approach defines weights according to the posterior probability of each candidate model. There are other methods that consider averaging based on selection criteria (e.g., \Citealp{burnham2003model, chen2012geostatistical}, among others). However, all of these approaches would also include poor performing models in their averages, where sparsity inducing priors have the ability to remove these models.

An important issue with the use of selection criteria is that sampling variability in the selected model is not incorporated directly into the selected model. For example, consider the Akaike information criterion (AIC) and the Bayesian information criterion (BIC). The AIC selects the model that minimizes an approximated Kullback-Leibler divergence to the true data generating process (see discussion in \citealt{acquah2010comparison}). The BIC is designed to approximate a Bayes factor (see discussion in \citealt{acquah2010comparison}), and hence, is often used when there are random effects. The values of these criteria are functions of the dataset itself, and hence, has sampling variability. Thus, as new data are generated, the ``best" model may change. This is true for a majority of the selection criteria used in the literature. For example, \citet{vaida2005conditional} define a corrected version of the AIC, referred to as the conditional AIC, for linear mixed-effects models. The conditional AIC penalizes the training error using the effective degrees of freedom \citep{hodges2001counting}, which has sampling variability. Another useful criterion introduced in \citet{huang2007optimal} selects spatial models where the penalty is based on the generalized degrees of freedom \citep{ye1998measuring} which again has sampling variability. 

To address this sampling variability issue, we apply a version of \citet{yekutieli2012adjusted}'s method to a BMA in order to directly incorporate a criterion into a Bayesian model. All of the aforementioned criteria can be interpreted as a type of covariance penalized error (CPE), which is described in detail by \cite{efron2004estimation}. As such, we use this general expression when extending \citet{yekutieli2012adjusted}'s approach to incorporate a criterion into BMA. \citet{yekutieli2012adjusted}'s selection-adjusted Bayes inference method involves truncating the support of the data based on selected values. Thus, the CPE is not treated as a plug-in estimator, and is instead is used to constrain the support of the Bayesian hierarchical model. Our method involves truncating the data and parameter space based on a selection criterion, which incorporates the criterion directly into the model in a principled way (i.e., through the support of the statistical model). Consequently, we refer to our model as the truncated CPE model. In this manuscript we choose the CPE, however, our constrained Bayesian perspective is flexible enough to incorporate several other criteria. 

We provide a result that shows \textit{every} proper Bayesian model for normally distributed data can be expressed as a type of truncated CPE model. In particular, one can augment a Bayesian model for normally distributed data with a uniformly distributed random variable (in a manner similar to \citet{damlen1999gibbs}) so that the posterior distribution can be expressed as a truncated CPE model. In our method, we explicitly make the truncation tighter, which can lead to better predictive performance. That is, we analytically show that the combination of \citet{yekutieli2012adjusted}'s method with BMA leads to better predictions in terms of mean squared error than BMA. This is particularly exciting because this is true even when $B=1$ (See right panel of Figure \ref{circle1}). Thus, we can improve upon a preferred single model as long as the truncating event is admissible. The size of the truncating event also has important practical implications. In particular, we can compare models through acceptance rates when implementing a Gibbs sampler, where we reject when the CPE is ``too large". That is, one model may reject more parameter values than another because its parameter space implies large values of CPE.


The remainder of this paper is organized as follows. In Section 2, we introduce the truncated CPE model, and in Section 3, we provide theoretical support. In particular, we show that every Bayesian model for normally distributed data can be interpreted as a truncated CPE model, and our specifications can lead to higher predictive performance than BMA in terms of mean squared errors. We illustrate this through a simulation study in Section 4. In Section 5, we analyze ACS period estimates over census tracts in central Missouri. Here, we apply our approach to space-time change of support \citep{bradley2015spatio}. This example demonstrates a case where only a single $B=1$ candidate Bayesian model is available, and that one can obtain better out-of-sample performances using the proposed truncated CPE model. We end with a discussion in Section 6. For ease of exposition, proofs are provided in the Appendix.

\section{Methodology}\label{sec:metho} 


\subsection{A Review of Prediction Error Estimation Method}
\label{sec:2.1}
Denote the observed data with \ser{Z}{n} and the $n$-dimensional observed data vector with $\bdm{z}\equiv(Z_{1},Z_{2},...,Z_{n})'$. We assume that \ser{Z}{n} are noisy representations of a subset of the latent random variables \ser{Y}{N} ($N\ge n$), and set $\bdm{y}\equiv(Y_{1},Y_{2},...,Y_{N})'$.
Specifically, we assume the following additive model 
\begin{equation}
	\label{equation:1}
	Z_{i} = Y_{i}+\epsilon_{i};\hspace{5pt} i=1,2,\dots,n,
\end{equation}
where the $\epsilon_{i}$'s are normal, mean zero, variance $\sigma^2>0$, are independent of $\epsilon_j$ and $Y_k$ for $j\neq i$, and $k=1,...,n.$ Now, suppose there are $B$ candidate models to predict $\bdm{y}$. These models all result in different predictors for $\bdm{y}$, which we denote with $\hat{\bdm{y}}_{b}: {\rm I\!R}^{n} \rightarrow {\rm I\!R}^{N};\ b=1,2,...,B$. For example, $\hat{\bdm{y}}_{b}\equiv(\hat{Y}_{1b},...,\hat{Y}_{Nb})'$ may be the posterior mean of $\bdm{y}$ using model $b$.

Let $Err_{i}(\bdm{z}, Y_i,\hat{Y}_{ib},\sigma^2)\equiv E[(Z_{i}^{(0)}-\hat{Y}_{ib})^2|\bdm{z}]=(Y_{i}-\hat{Y}_{ib})^2+\sigma^2$ be the prediction error, where $Z_{i}^{(0)}$ is an independent replicate of $Z_{i}$. The term $E[Err_{i}(\bdm{z}, Y_i,\hat{Y}_{ib},\sigma^2)]$ is not observed. In practice, one can more easily compute the training error, $err_{i}(Z_i,\hat{Y}_{ib})\equiv(Z_{i}-\hat{Y}_{ib})^2$. \citet{efron1983estimating,efron1986biased,efron2004estimation} derived an important expression of the prediction error, 
	\begin{equation}
		\label{equation:2}
		E[Err_{i}(\bdm{z}, Y_i,\hat{Y}_{ib},\sigma^2)] = E[err_{i}(Z_i,\hat{Y}_{ib})+2cov(\hat{Y}_{ib},Z_{i})];\hspace{5pt} i=1,\dots,n,\ b=1,\dots,B,
	\end{equation}
where the expectation is taken with respect to $\bdm{z}|\bdm{y},\sigma^2.$ Equation (\ref{equation:2}) shows that $err_{i}$ is biased for $E[Err_{i}(\bdm{z}, Y_i,\hat{Y}_{ib},\sigma^2)]$, which leads to the following criterion referred to as the CPE,
\begin{equation}
    \label{equation:cpe}
	 CPE(\bdm{z},\hat{\bdm{y}}_{b})=\sum_{i=1}^{n}err_i(Z_i,\hat{Y}_{ib})+2\sum_{i=1}^{n}cov(\hat{Y}_{ib}, Z_{i});\hspace{5pt} b=1,\dots,B,
\end{equation} which is unbiased for 
\begin{equation*}
 \sum_{i=1}^{n}E(Err_i(\bdm{z}, Y_i,\hat{Y}_{ib},\sigma^2))=\sum_{i=1}^{n}E(Y_i-\hat{Y}_{ib})^2+n\sigma^2.
\end{equation*}
These fundamental results show that the training error, $err_{i}$ needs to be corrected by a penalty (i.e., a covariance, hence the name CPE) to be an unbiased estimation for $E[Err_{i}]$. This CPE criterion is  well-known to be a general expression of several criteria introduced in the literature. For example, the AIC, Mallow's $C_{p}$ \citep{mallows1973some}, and Stein's unbiased risk estimator \citep{stein1981estimation} are all special cases of the CPE (see \Citealp{efron2004estimation} for a discussion). 

This criterion, while very useful, has a limitation that we focus on in this paper. Namely, the CPE is a statistic (more formally a method of moments estimate of $\sum_{i}E[Err_{i}]$), and hence has sampling variability. This sampling variability can have an effect on the chosen models. Consider the following simulated example to illustrate the issue of sampling variability in selection criterion:

\begin{itemize}
	\item Simulate 1000 replicate with $n=N=200$.
	\item  Consider a multiple regression model with $\bdm{x}_1, \bdm{x}_2, \bdm{x}_3$, each a 200-dimensional vector, where the elements are chosen independently from a standard normal distribution.
	\item Let the $200\times4$ matrix $\bdm{X}_{b}=[\bdm{1}_{200},\bdm{x}_1\delta_{1}, \bdm{x}_2\delta_{2}, \bdm{x}_3\delta_{3}]=(\bdm{x}_{1b}',\dots,\bdm{x}_{200b}')'$, where $\delta_{i}$ is either zero or one, $\bdm{1}_{200}$ is a 200-dimensional vector of ones, define $\bdm{y}=\bdm{X}_{b}{\bm\beta}$, where the value of $\bm{\beta}=(\beta_0,\beta_1,\beta_2,\beta_3)'$ is arbitrarily chosen to be $(2,1,1,0)'$
	\item For a given 200-dimensional data vector $\bdm{z}$, we consider implementing the following models for $\bdm{y}$:
	\begin{align}
	 \label{equation:long}
        \nonumber
		&I(b=1)= I(\delta_1=\delta_2=\delta_3=0)\\
		\nonumber
		&I(b=2) = I(\delta_1=1,\delta_2=\delta_3=0)\\
		\nonumber
		&I(b=3) = I( \delta_1=\delta_3=0, \delta_2=1)\\
		\nonumber
		&I(b=4) = I( \delta_1=\delta_2=0, \delta_3=1)\\		
		&I(b=5) = I(	\delta_1=\delta_2=1,\delta_3=0)\\ 
		\nonumber
		&I(b=6) = I(\delta_1=\delta_3=1,\delta_2=0)\\
		\nonumber
		&I(b=7) = I(\delta_2=\delta_3=1,\delta_1=0)\\
		\nonumber
		&I(b=8) = I( 
		\delta_1=\delta_2=\delta_3=1),
	\end{align}
\end{itemize}
\noindent
where $I(\cdot)$ is an indicator function. Then let $\hat{\bdm{y}}_b$ be the ordinary least squares estimator with eight different choices of covariates based on (\ref{equation:long}). The Mallow's $C_p$ is given by 
$$C_p(\bdm{z},\hat{\bdm{y}}_b)=\sum_{i=1}^{200}err_i(Z_i,\hat{Y}_{ib}) +2\sigma^2p(b); \hspace{5pt} b=1,\dots,8,$$
where $p(b)$ is the number of non-zero regression coefficients identified in the model $b$. Note that for
$$ \hat{\bdm{y}}_b=\bdm{X}_{b}(\bdm{X}_{b}'\bdm{X}_{b})^{-1}\bdm{X}_{b}'\bdm{z},$$ we have the covariance in Equation (\ref{equation:cpe}) is given by $$ \sum_{i=1}^{200}cov(\bdm{x}_{ib}'(\bdm{X}_{b}'\bdm{X}_{b})^{-1}\bdm{X}_{b}'\bdm{z},Z_i)=trace(\bdm{X}_{b}(\bdm{X}_{b}'\bdm{X}_{b})^{-1}\bdm{X}_{b}')\sigma^2=p(b)\sigma^2=\sigma^2b, $$
which shows that Mallow's $C_p$ is a special case of the CPE when selecting covariates using the ordinary least squares \citep[e.g.,see][among others]{efron2004estimation}. Then denote the selected model with 
$$ \hat{b}=arg\min_{b=1,\dots,8}C_p(b). $$

\begin{table}[t]
	\centering
	\caption{The proportion of times $\hat{b}=b$ by $\sigma$ over 1000 independent replicates of the vector $\bdm{z}$.}

	\begin{tabular}{lrrrrrrrr}
		&  &  &  & b &  &  &  & \\
		\hline
		&1 & 2 &  3 & 4  & 5  & 6 & 7 & 8 \\ \hline
	$\sigma=0.5$& 0 & 0 & 0& 0&83.5\% &0  &0 &16.5\% \\
	$\sigma=1$  & 0 & 0 & 0& 0  &85.2\% &0  &0 &14.8\%  \\
	$\sigma=2$  & 0 & 0 & 0& 0  &83.0\% &0  &0 &17.0\%  \\
	$\sigma=3.5$&0 & 0.4\% & 0.4\%& 0  & 82.3\%&0  &0 &16.9\%  \\ \hline
	\end{tabular}
	
	\label{table:cpe_eg} 
\end{table}
\noindent
From Table \ref{table:cpe_eg},  we present the proportion of times $\hat{b}=b$ by $\sigma$ over 1000 independent replicates of the vector $\bdm{z}$. For each $\sigma$, $83\%$ of the time we roughly select the correct value of $b=5$,
but we consistently (over $\sigma$) select the incorrect full model around $17\%$ of time. This is consistent with the literature, where several (but not all) selection criteria tend to select more complicated models \citep{rao1989strongly, maraun2018statistical}. This also demonstrates the weakness of selection criteria discussed in the Introduction. That is, high sampling variability in $C_p$ can lead to incorrectly chosen models.
\subsection{A Review of Bayesian Model Averaging (BMA)}
Bayesian model averaging addresses model uncertainty (as demonstrated in Table \ref{table:cpe_eg}) by directly modeling $b$ with a prior distribution. Let $\pi(b)$ be the prior mass for model $b$ such that $\sum_{j=1}^{B}\pi(b=j)=1$. Under BMA, inference on the quantity of interest (here is $\bdm{y}$), can be obtained through the probability density function (pdf) of $\bdm{y|z}$. This can be computed with 
\begin{equation}
\pi(\bdm{y|z})= \sum_{j=1}^{B}\pi(\bdm{y}|\bdm{z},b=j)\pi(b=j|\bdm{z}),
\end{equation}
which is a weighted average of the distribution of $\bdm{y}$ given each model and data, and the weights are posterior probability of the model. 
The choice of prior specifications for the candidate models have an important impact in practice. Let's revisit the small simulation example in Section \ref{sec:2.1}, where notice $b = 1,2,3,4,6,7$ in Table \ref{table:cpe_eg} were nearly never selected using Mallow's $C_p$. This leads us to consider the case where $\pi(b=j) = 1/8$ (for all $j$) and the case 
\begin{equation}
\label{equation:4}
\pi(b=j)=
\begin{cases}
1/2& \text{j = 5,8}\\
0& \text{otherwise}.
\end{cases}
\end{equation}
Consider the case $\sigma=2$. Figure \ref{BMA} contains a histogram of, 
$\sum_{i=1}^{n}(Y_i-\hat{Y}_{iv})^2
  - \sum_{i=1}^{n}(Y_i-\hat{Y}_{iw})^2
$,
where $\hat{Y}_{iv}$ is the posterior mean using $\pi(b=j) = 1/8,$ and $\hat{Y}_{iw}$ is the posterior mean using Equation (\ref{equation:4}). The majority of values in Figure \ref{BMA} are consistently positive, which suggests better predictions when using $\pi(b=j)$ in (\ref{equation:4}).
Here, we can see that the choice of prior on the models have a clear impact by informally using Table \ref{table:cpe_eg} (or the $CPE$) to reduce the parameter space (of $b$). The improvements, by using (\ref{equation:4}), are not surprising. Poor performing values in the parameter space are averaged in BMA when $\pi(b=j)=1/8,$ that are not averaged when using $\pi(b=j)$ in (\ref{equation:4}). The prior distribution $\pi(b)$ in (\ref{equation:4}) informally incorporates $CPE$, which was formed via Table \ref{table:cpe_eg}, but does not account for the sampling variability of $CPE$. Thus, our goal is to formally incorporate $CPE$, by accounting for the variability of $CPE$.
\begin{figure}[H]
	\centering
	\includegraphics[width=0.8\linewidth]{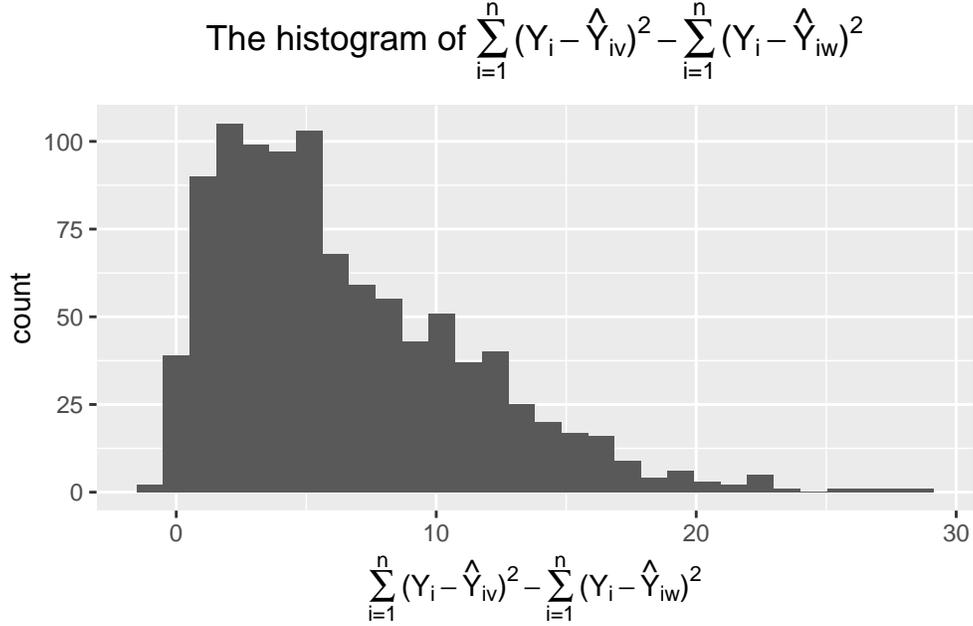} 
	\caption{The histogram of $\sum_{i=1}^{n}(Y_i-\hat{Y}_{iv})^2
		- \sum_{i=1}^{n}(Y_i-\hat{Y}_{iw})^2$ (i.e. the difference in squared error) by $\sigma=2$ over 1000 independent replicates of the vector $\bdm{z}$.}
	\label{BMA}
\end{figure}

\subsection{The Proposed Model}
 The statistical model we use for inference is defined as the product of the following conditional and marginal probability density functions: 
\begin{equation}
\label{equation:5}
	\pi(\bdm{z},\bdm{y},\bm{\theta},b|\kappa)\propto f(\bdm{z}|\bdm{y},\sigma^2)\pi(\bdm{y}|\bm{\theta},b)
	\pi(\bm{\theta})\pi(b)I\{\widehat{Err}<\kappa\}; b=1,\dots,B,
\end{equation}
 \noindent
where $I(\cdot)$ is the indicator function, $f(\bdm{z}|\bdm{y},\sigma^2)$ is the normal distribution with mean $\bdm{y}$ and constant variance $\sigma^2>0$, $\bm{\theta}$ is the generic real-valued parameter vector, $\pi(\bdm{y}|\bm{\theta},b)$ is the process model, $\pi(\bm{\theta})$ is the prior for $\bm{\theta}$, $\pi(b)$ is the prior probability of the model $b$, the value of $\kappa>0$ is a pre-specified real value and is crucial for our model (see Section 3, and 4 for more discussion), and $\widehat{Err}$ is an {unbiased} estimator for $Err$ (e.g., CPE). In our general expression of the model in (\ref{equation:5}), we allow for several estimates of $Err$, where besides CPE, one might use an information criterion or cross-validation. The model in Equation (\ref{equation:5}) allows for many special cases. For example, in our application $B=1$, and we show that (\ref{equation:5}) can lead to improvements in a single model.

When using CPE to estimate $Err$, we introduced $\hat{\bdm{y}}$ into our notation for $CPE(\bdm{z},\hat{\bdm{y}}(\bm{\theta},b)),$ where $\hat{\bdm{y}}(\bm{\theta},b)$ is a generic predictor of $\bdm{y}$. We also introduce the possible functional dependence on $\bm{\theta}$ and $b$ into our notation for $\hat{\bdm{y}}(\bm{\theta},b)$. This strategy is inspired by \cite{yekutieli2012adjusted}'s method. His selection-adjusted Bayes inference method involves truncating the support of the data. Our method differs because it involves truncating the support based on CPE.  The $CPE(\bdm{z},\hat{\bdm{y}}(\bm{\theta},b))$ is directly incorporated into the model through $I\{CPE(\bdm{z},\hat{\bdm{y}}(\bm{\theta},b))<\kappa\},$ and hence, does not have unaccounted for variability in Equation (\ref{equation:5}). Specifically, we mean that the joint posterior distribution of our model is given by 
\begin{equation}
  \label{equation:6}
  \frac{f(\bdm{z}|\bdm{y},\sigma^2)\pi(\bdm{y}|\bm{\theta},j)
  	\pi(\bm{\theta})\pi(j)I\{CPE(\bdm{z},\hat{\bdm{y}}(\bm{\theta},j))<\kappa\}}
  {\sum_{b=1}^{B}\hspace{5pt}\underset{CPE(\bdm{z},\hat{\bdm{y}}(\bm{\theta},b))<\kappa}{\int\int}f(\bdm{z}|\bdm{y},\sigma^2)\pi(\bdm{y}|\bm{\theta},b)\pi(\bm{\theta})\pi(b)I\{CPE(\bdm{z},\hat{\bdm{y}}(\bm{\theta},b))<\kappa\}d\bdm{y}d\bm{\theta}},
\end{equation}
\noindent
 for $j=1,\dots,B$, which does not treat $CPE(\bdm{z},\hat{\bdm{y}}(\bm{\theta},b))$ as a plug-in estimator (causing unaccounted for variability), but rather uses CPE to constrain the support of the Bayesian hierarchical model. Equation (\ref{equation:6}) is well defined provided that $\kappa$ is not specified so small that the integral is equal to zero. More empirically motivated discussions on the choice of $\kappa$ are given by Section 4 and 5.

The joint posterior distribution in Equation (\ref{equation:6}) shows explicitly how we combine BMA, classical model selection,  and criteria. Specifically, a prior is placed on the model $b$, and the parameter space of this model is constrained to a ``good predictive set" by using $I\{CPE(\bdm{z},\hat{\bdm{y}}(\bm{\theta},b))<\kappa\}$. That is, for example in Figure \ref{circle1}, $I\{CPE(\bdm{z},\hat{\bdm{y}}(\bm{\theta},b))<\kappa\}$ subsets the three models represented by circles to the black region.

Besides CPE, one can also characterize  the prediction capacity of a model by means of nonparametric methods. Cross-validation (CV) is a popular nonparametric technique to assess the prediction ability of a model \citep{efron2004estimation}. There are several types of CV, including K-fold CV (KFCV) and leave-one out CV (LOOCV) \citep{geisser1975predictive}. For KFCV, one randomly splits the data into $K$ approximately equal subgroups, or folds. Each fold is successively treated as a validation set, and the rest of the $K-1$ folds are used to train model to produce a predictor at the validation set. Then $Err$ is estimated by the average of the squared difference between each validation set and its associated predictor. {LOOCV is a special case of KFCV, where $K$ equals to the size of the data set. Each data point is consecutively used for validation, and the remaining parts of the data set account for training model, and accordingly obtaining a predictor at the validation set.  }
\section{Theoretical Justification}
\subsection{Motivation}
A key point that motivates the truncated CPE model in (\ref{equation:5}) is that \textit{every} proper Bayesian hierarchical model for normally distributed data can be interpreted as a type of truncated CPE model. Specifically, one can augment any proper Bayesian hierarchical model, using the technique introduced in \cite{damlen1999gibbs}, so that the CPE is bounded above. We formally state this result below in Theorem 1.\\

\begin{theorem}
	\label{Them:3} 
	Suppose
	 $\pi(\bdm{z}|\bdm{y},\bm{\theta},\sigma^2,b)=f(\bdm{z}\vert\bdm{y},\sigma^2)h(\bdm{z},\bm{\theta},b)$, $\pi(b)>0$, $\sum_{b}\pi(b) = 1$, $\pi(\textbf{y}\vert \bm{\theta},b)$ and $\pi(\bm{\theta})$ are proper densities, recall $f(\bdm{z}\vert\bdm{y},\sigma^2)$ is the multivariate normal distribution with mean $\textbf{y}$ and constant known variance $\sigma^{2}$, and $h(\bdm{z},\bm{\theta},b)$ is a non-negative real-valued function such that $0<\int f(\bdm{z}\vert\bdm{y},\sigma^2)h(\bdm{z},\bm{\theta},b)d\bdm{z}<\infty$. Then, for $u$ uniformly distributed on $(0,1)$ and $r =\frac{CPE(\bdm{z},\hat{\bdm{y}})}{2log\left(f(\bdm{z}\vert \bdm{y},\sigma^{2})\right)} + 1$, we have that the posterior distribution
	\begin{equation*}
	\pi(\bdm{y},\bm{\theta},b\vert \bdm{z}) = \frac{1}{\pi(\bdm{z})}\int_{0}^{1} \pi(\bdm{y},\bm{\theta},b,\bdm{z}\vert u) \pi(u) du,
	\end{equation*}
	where $\pi(\bdm{z})$ is the density for the marginal distribution of the data,
	\begin{equation}
	\label{equation: Rlship}
		\pi(\bdm{z},\bdm{y},\bm{\theta},b\vert u) =
		f(\bdm{z}\mid \bdm{y},\sigma^2)^{r}\pi(\bdm{y}\vert \bm{\theta},b) \pi(\bm{\theta})\pi(b)
		I\{CPE(\bdm{z},\hat{\bdm{y}})<\kappa^*\}h(\bdm{z},\bm{\theta},b),		
	\end{equation}
	and $\kappa^*=-2log(u)$.
\end{theorem}
\noindent
$Proof:$ See Appendix A.\\

\noindent
When $h(\bdm{z},\bm{\theta},b)\equiv 1$ then Equation (\ref{equation: Rlship}) in Theorem 1 shows that any generic Bayesian hierarchical model for normally distributed data (with constant variance) is a truncated CPE model, where the CPE is truncated above by $\kappa^{*}$ in (\ref{equation: Rlship}). That is, Equation (\ref{equation: Rlship}) with $h(\bdm{z},\bm{\theta},b)\equiv 1$ is directly analogous to the truncated CPE model in (\ref{equation:5}). 

Setting $h(\bdm{z},\bm{\theta},b) = I\{CPE(\bdm{z}),\hat{\bdm{y}}(\bm{\theta},b)<\kappa\}$ in Theorem 1 implies that the proposed truncated CPE model in (\ref{equation:5}) truncates the CPE above by $min(\kappa^{*},\kappa)$, since the product
\begin{equation*} I\{CPE(\bdm{z},\hat{\bdm{y}}(\bm{\theta},b))<\kappa^{*}\}I\{CPE(\bdm{z},\hat{\bdm{y}}(\bm{\theta},b))<\kappa\} = I\{CPE(\bdm{z},\hat{\bdm{y}}(\bm{\theta},b))<min(\kappa^{*},\kappa)\}. 
\end{equation*}
\noindent
Thus, one can interpret our truncated CPE model in (\ref{equation:5}) as a minor modification to any Bayesian hierarchical model, where one replaces the implicit bound on the CPE (i.e., $\kappa^{*}$) with $min(\kappa^{*}, \kappa)$. Changing $\kappa^{*}$ to $min(\kappa^{*}, \kappa)$ has two important consequences. First, changing $\kappa^{*}$ (or $h(\bdm{z},\bm{\theta},b)\equiv 1$) to $min(\kappa^{*}, \kappa)$  (or $h(\bdm{z},\bm{\theta},b) = I(CPE(\bdm{z},\hat{\bdm{y}}(\bm{\theta},b))<\kappa)$) changes the data model from a normal distribution to a type of truncated normal distribution. However, as shown in Theorem 1 the implied posterior for either choice of data model (truncated or un-truncated) stays the same (i.e., Equation (\ref{equation: Rlship}) and (\ref{equation:5}) are analogous). Furthermore, changing the distribution of the data is reasonable in our model selection setting, as we are allowing for the possibility of model mis-specification. Second, changing $\kappa^{*}$ to $min(\kappa^{*}, \kappa)$ can lead to smaller mean squared prediction errors, which we discuss in detail in the subsequent Section 3.2.

\subsection{Mean Squared Prediction Error Properties}
Constraining a Bayesian hierarchical model based on the CPE implicitly constrains the unobserved $E\{\sum_{i=1}^{n}(Y_i-\hat{Y}_i)^2\}$. To investigate this consider the setting where the predictor $\hat{\bdm{y}}$ is specified to be the Best Linear Unbiased Prediction (BLUP) \citep{ravishanker2020first}, $$\hat{\bdm{y}}(\bm{\theta},b)= \bdm{\mu}_Y(\bm{\theta},b)+\bm{\Sigma}_Y(\bm{\theta},b)\bm{\Sigma}_Z^{-1}(\bm{\theta},b)\{\bdm{z}-\bdm{\mu}_Y(\bm{\theta},b)\},$$ where $\bdm{\mu}_Y(\bm{\theta},b)$ is the mean of the process model $\pi(\bdm{y}|\bm{\theta},b),$ $\bm{\Sigma}_Y(\bm{\theta},b)$ is the process model's covariance, and $\bm{\Sigma}_Z(\bm{\theta},b)$ is the covariance of $\bdm{z}$ from $f(\bdm{z}|\sigma^2,b,{\bm{\theta}})=\int f(\bdm{z}|\bdm{y},\sigma^2)\pi(\bdm{y}|\bm{\theta},b)d\bdm{y}.$ The CPE for this specification of $\hat{\bdm{y}}$ is computed using $\hat{\bdm{y}}(\bm{\theta},b)$ as follows \citep{efron2004estimation}:
$$CPE(\bdm{z},\hat{\bdm{y}}(\bm{\theta},b))=\{\bdm{z}-\hat{\bdm{y}}(\bm{\theta},b)\}'\{\bdm{z}-\hat{\bdm{y}}(\bm{\theta},b)\}+2\sigma^2trace\{\bm{\Sigma}_Y\bm{\Sigma}_Z^{-1}\}.$$ 
\noindent
where the penalty term is referred to as the effective degrees of freedom \citep{hodges2013richly}. Then the following result shows that $\kappa$ can be chosen in a manner that leads to smaller mean squared prediction error. 
 \begin{theorem}
 	\label{Them:1}
 	 Assume $\bdm{z}|\bdm{y},\sigma^2\sim N(\bdm{y},\sigma^2\textbf{I}),$ and let $\hat{\bdm{y}}_{tc}=(\hat{Y}_{1,tc},...,\hat{Y}_{n,tc})'$ be the element-wise posterior median of $\hat{\bdm{y}}$ using the model in (\ref{equation:6}). Then,
 	  \begin{equation}
 	  \label{equation:7}
 	  	E\{\sum_{i=1}^{n}(Y_i-\hat{Y}_{i,tc})^2\}< E\{\sum_{i=1}^{n}(Y_i-\hat{Y}_{i,m})^2\},	
 	  \end{equation}
where $\hat{Y}_{i,m}$ is a generic real-valued predictor of $Y_i$, and $\kappa= E\{\sum_{i=1}^{n}(Y_i-\hat{Y}_{i,m})^2\}+n\sigma ^2$. We assume this choice of $\kappa$ produces a model in (\ref{equation:6}) that is proper.
 \end{theorem}
\noindent
$Proof:$ See Appendix A.

Theorem \ref{Them:1} shows that this use of CPE can be used to improve the unobserved MSPE of any predictor $\hat{Y}_{i,m}$ given the conditions in Theorem \ref{Them:1}. This is true despite the fact that CPE has sampling error. The result is general because the assumptions on $\bdm{y}$ are only used to define the BLUP and are not used to obtain (\ref{equation:7}). Also, Theorem 1 shows that on average our model is restricted to a good predictive performing set, where`` good predictive performance" is defined as $CPE(\bdm{z},\hat{\bdm{y}})<E\{\sum_{i=1}^{n}(Y_i-\hat{Y}_{i,m})^2\}+n\sigma ^2.$ This result cannot be directly used in practice since for the term, $E\{\sum_{i=1}^{n}(Y_i-\hat{Y}_{i,m})^2\}+n\sigma ^2$, the expectation is taken with respect to $f(\bdm{z}|\bdm{y},\sigma^2)$ and $\bdm{y}$ and the true model for $\bdm{y}$ is assumed unknown. However, Theorem \ref{Them:1} does suggest a choice of $\kappa$ exists that can lead to a good predictive results. As a result, in practice several values of $\kappa$ are considered, where small values would imply a better prediction error. However, one should keep in mind the admissibility of the set $\{CPE[\bdm{z},\hat{\bdm{y}}(\bm{\theta},b)]<\kappa\}$ when choosing small values of $\kappa$ (e.g., $\kappa=0$ would be inadmissible).

Theorem \ref{Them:1} can be extended from a multivariate vector to a random process. To do this, we introduce notation that treats $Z,$ $Y$ and $\epsilon$ as processes: $ Z(\bdm{s})=Y(\bdm{s})+\epsilon(\bdm{s}),$ where $Z(\bdm{s})$ is the noisy version of the latent process $Y(\bdm{s})$ at location $\bdm{s}\in D \subset \rm I\!R^d $, $\epsilon(\bdm{s})$ is normally distributed with mean zero, constant variance $\sigma ^2>0$, and $\epsilon(\bdm{s}_i)$ is independent of $\epsilon(\bdm{s}_j)$ for $i\neq j$ and $\bdm{s}_i,\bdm{s}_j \in D$. Let $\bdm{z}=(Z(\bdm{s}_1),...,Z(\bdm{s}_n))', \bdm{y}=(Y(\bdm{s}_1),...,Y(\bdm{s}_n))'$, where $\bdm{s}_1,...,\bdm{s}_n$ are locations associated with the observed data. Then, the model in (\ref{equation:5}) stays the same. For $\bdm{s}_0\in D,$  define the Kriging Predictor \citep{cressie1993spatial} as $\hat{Y}(\bdm{s}_0)=  
\bdm{\mu}(\bdm{s}_0)+cov\{Y(\bdm{s}_0),\bdm{z}\}\bm{\Sigma}_Z^{-1}\{\bdm{z}-\bdm{\mu}_Y\}$.
 \begin{corollary}
	\label{Them:2}
	Let $D$ be a spatial domain and $f(\bdm{\cdot}): D\rightarrow \rm I\!R$ be an intensity function. Suppose we observe normal data $Z(\textbf{s}_{i})$ with mean $Y(\bdm{s}_{i})$ and variance $\sigma^{2}>0$ for $i = 1,\ldots, n$.
	The notation $\hat{Y}_{tc}(\bdm{s})$ represents the posterior median of the Kriging predictor $Y(\bdm{s})$ and $\hat{Y}_{m}(\bdm{s})$ be a generic real-value predictor of $Y(\bdm{s}).$ Let $\textbf{s}_{1},\ldots, \textbf{s}_{n}$ be independent and identically distributed according to $f(\textbf{s})$. Then, as $n\rightarrow \infty$,	 
	\begin{equation}
	\label{equation:8}
\underset{D}{\int}E\{Y(\bdm{s})-\hat{Y}_{tc}(\bdm{s})\}^2f(\bdm{s})d\bdm{s}< \underset{D}{\int}E\{Y(\bdm{s})-\hat{Y}_{m}(\bdm{s})\}^2f(\bdm{s})d\bdm{s},	
	\end{equation}
where $\kappa=\sum_{i = 1}^{n}E\{Y(\bdm{s}_{i})-\hat{Y}_{m}(\bdm{s}_{i})\}^2 + n\sigma^{2}.$ We are assuming that this choice of $\kappa$ leads to a proper model in (\ref{equation:6}).
\end{corollary}
 \noindent
$Proof:$ See Appendix A.

Again in practice, we do not know the value of $\kappa=\sum_{i = 1}^{n}E\{Y(\bdm{s}_{i})-\hat{Y}_{m}(\bdm{s}_{i})\}^2 + n\sigma^{2}$. However, Corollary \ref{Them:2} shows a value of $\kappa$ exists, where we can obtain improvements on $\hat{Y}_{m}(\cdot).$ We empirically investigate the choice of $\kappa$ in practice.

\section{Simulation Study}
In this section, we perform an ``empirical simulation study." By this, we mean the data generating mechanism is calibrated towards the dataset. This strategy is done in an effort to produce a realistic simulated dataset that differs from the model we fit. This aids in producing realistic simulated data and assessing departures from model assumptions. Thus, we generate data from the following statistical model:

\begin{equation}
	\bdm{z}\sim N(\bdm{L},\sigma^2\bdm{I}_n),
\end{equation} 
\noindent
where $\bdm{I}_n$ is an $n\times n$ identify matrix and $n=112$.

Let $\bdm{L}=(L_1,L_2,...,L_n)'$ be a $n$-dimensional dataset (\url{www.biostat.umn.edu/~brad/data2.html}, \cite{}) consisting of the log thickness of radioactive materials at each of $n=112$ sites contained within the Radioactive Waste Management Complex region associated with the Idaho National Engineering and Environmental Laboratory. We use covariates 
A-B Elevation, and Surf Elevation. The value of $\sigma^2$ is chosen in a way that controls the signal to noise ratio (SNR). Specifically, we choose $SNR$ and solve for $\sigma^2$ in $SNR=\frac{1}{111}\sum\limits_{i=1}^{112}(L_i-\bar{L})^2/\sigma^2$, where $\bar{L}=\frac{1}{112}\sum\limits_{i=1}^{112}L_i$. We give our choices for $SNR$ when describing our analysis of variance (ANOVA) later in this section.

The  model we fit to the simulated data is a Bayesian hierarchical model with truncated data model:
\begin{align}
\label{simu:model}
\nonumber
&\text{Data Model:	}
\bdm{z}|\bdm{y},\bm{\theta}, b\sim{N} (\bdm{y},\sigma^2\bdm{I}_n)I\{CPE[\bdm{z},\hat{\bdm{y}}(\bm{\theta},b)]<\kappa\}    \\
\nonumber
&\text{Process Model:	}\bdm{y}|\bm{\theta},b,\sim N(\bdm{X}\bm{\beta},\bdm{C}(b,\tau^2))\\
&\text{Parameter Model 1:	}\bm{\beta}\sim 
N(\bdm{0}_p,10\bdm{I}_p)\\
\nonumber
&\text{Parameter Model 2:	}\tau^2\sim IG(1,0.01)		
\\
\nonumber										
&\text{Parameter Model 3:	}
\pi(b=j)=\frac{1}{6};\hspace{20pt} j=10,15,...,35,
\end{align}
\noindent
where $IG(\cdot)$ is the inverse gamma distribution, $\bdm{X}$ is a $n\times p$ matrix, $p=3$ since we take the intercept and the aforementioned 2 covariates into consideration, and $\bm{\beta}$ are the associated coefficients. The $(i,j)$-th element of $n\times n$ matrix $\bdm{C}(b,\tau^2)$ is specified as $\tau^2exp(-b||\bdm{s}_i-\bdm{s}_j||)$, $\sigma^2>0$ is assumed as a known value, $||\bdm{s}_i-\bdm{s}_j||$ is the Euclidean distance between the $i$-th and $j$-th location,  $\bdm{0}_n$ is a $n$-dimensional zero vector, and  $\bm\theta=(\bm{\beta}',\tau^2)'$. In Appendix B, we derive the full-conditional distributions associated with this model.

We consider two crucial factors that influence $\hat{\bdm{y}}$ of them and specify their levels for an analysis of variance (ANOVA) as follows: $SNR$ with 3 levels, $SNR=3,5,10 $; the values for $\kappa$ are set equal to the $d$-th percentile of the set $\{CPE[\bdm{z},\hat{\bdm{y}}(\bm\theta^{[1]},b^{[1]})],\dots,CPE[\bdm{z},\hat{\bdm{y}}(\bm\theta^{[G]},b^{[G]})]\}$ for levels $d=0.1, 0.5, 0.9$, where $\bm\theta^{[i]}$, $b^{[i]}$ are the $i$-th Markov Chain Monte Carlo (MCMC) replicate for $\bm\theta$ and $b$ respectively, and the $CPE[\bdm{z},\hat{\bdm{y}}(\bm\theta,b)]$ is the $CPE$ from the untruncated model; $G$ is the length of the MCMC.  We simulate $100$ independent replicates of the data vector $\bdm{z}$ and implement our model as well as BMA, both of which are computed using a Gibbs sampler (see Appendix B). We evaluate the models using the sum of squared residuals, and we define as ``Response" in our ANOVA, whose form is  $\sum_{i=1}^{n}(Y_i-\hat{Y}_{i,tc})^2-\displaystyle \sum_{i=1}^{n}(Y_i-\hat{Y}_{i,m})^2$. We use a MCMC with length of $12,000$ and a burn-in of $2,000$ and use trace plots to assess convergence visually for a single replicate of the simulated $\bdm{z}$.

We analyze the effect of the aforementioned factors SNR and Percentile $d$ on the Response by using an ANOVA with $100$ independent replicates of the vector $\bdm{z}$ per factor level conbination. From Table \ref{table:anova}, we can see that the main effects and the interaction between them are highly significant. To visualize the main effects and the interaction, we use boxplots and an interaction plot.  In Figure \ref{plot:main_effect}, we see that as SNR increases, the boxplot for the Response shows less variability, but is centered below zero. Negative values suggest that the truncated model surpasses BMA in terms of squared errors. As Percentile $d$ increases, the boxplot for the Response is less negative for $d=0.1$ and $0.9$ than it is when $d=0.5$. From Figure \ref{plot:interaction}, it can be seen that the interaction is due to the fact that the slope of the line for $d=0.5$ is much steeper than the lines for $d=0.1$ and $0.9$. Also, the behavior when $d=0.9$ is very similar to that of $d=0.1.$ When $SNR=3$, BMA does not outperform our method when $d=0.1$ and $0.9$, and does considerably worse when $d=0.5$ in practice. These results conform to intuition. When $d$ approaches $1$, there should be no difference between the truncated model and BMA. Following our discussion after Theorem \ref{Them:1}, small values of $\kappa$ may imply inadmissibility, which violates the condition of our Thoerem.

	Based on above results, the values of Response (i.e., $ \sum_{i=1}^{n}(Y_i-\hat{Y}_{i,tc})^2  \textless \sum_{i=1}^{n}(Y_i-\hat{Y}_{i,m})^2$ ) for $d=0.5$ are uniformly less than zero. Thus, we suggest using $d=0.5$ in practice. When  $d$ is $0.1$ or $0.9$, the values of Response are less than zero, but still less preferable when it comes to sums of squared error as when $d=0.5$. In practice, one might use an information criterion to choose $\kappa$. Therefore, our method do as appear to improve the prediction accuracy with respect to the sum of squared residuals.
\begin{table}[t]
	\centering
	\caption{Two-way ANOVA table. The degrees of freedom (DF), sum of squares error (Sum Sq), mean squared error (Mean Sq), F statistics, and P-value are listed. We include two main effects, $SNR$ and $d$, and the interaction between $SNR$ and $d$.}
	
	\begin{tabular}{lrrrrr}
		\hline
		& \multicolumn{1}{c}{DF} & \multicolumn{1}{c}{Sum Sq} & \multicolumn{1}{c}{Mean Sq} & \multicolumn{1}{c}{F value} & \multicolumn{1}{c}{Pr(\textgreater{}F)}  \\ \hline
		SNR  & 2& 1.4417 & 0.72085 & 44.1791 &$<2.2\times10^{-16}$\\
		$d$&2 & 1.8038 & 0.90192 & 55.2765 
		&$<2.2\times10^{-16}$\\
		SNR:$d$& 4  & 0.3221       & 0.08052        & 4.9348        &$0.0006138$\\
		Residuals    & 891  & 14.5380      &  0.01632       &         & \\ \hline
	\end{tabular}

	\label{table:anova} 
\end{table}
\begin{figure}[H]
	\centering
	\begin{subfigure}{.50\textwidth}
		\centering
		\includegraphics[width=1\linewidth]{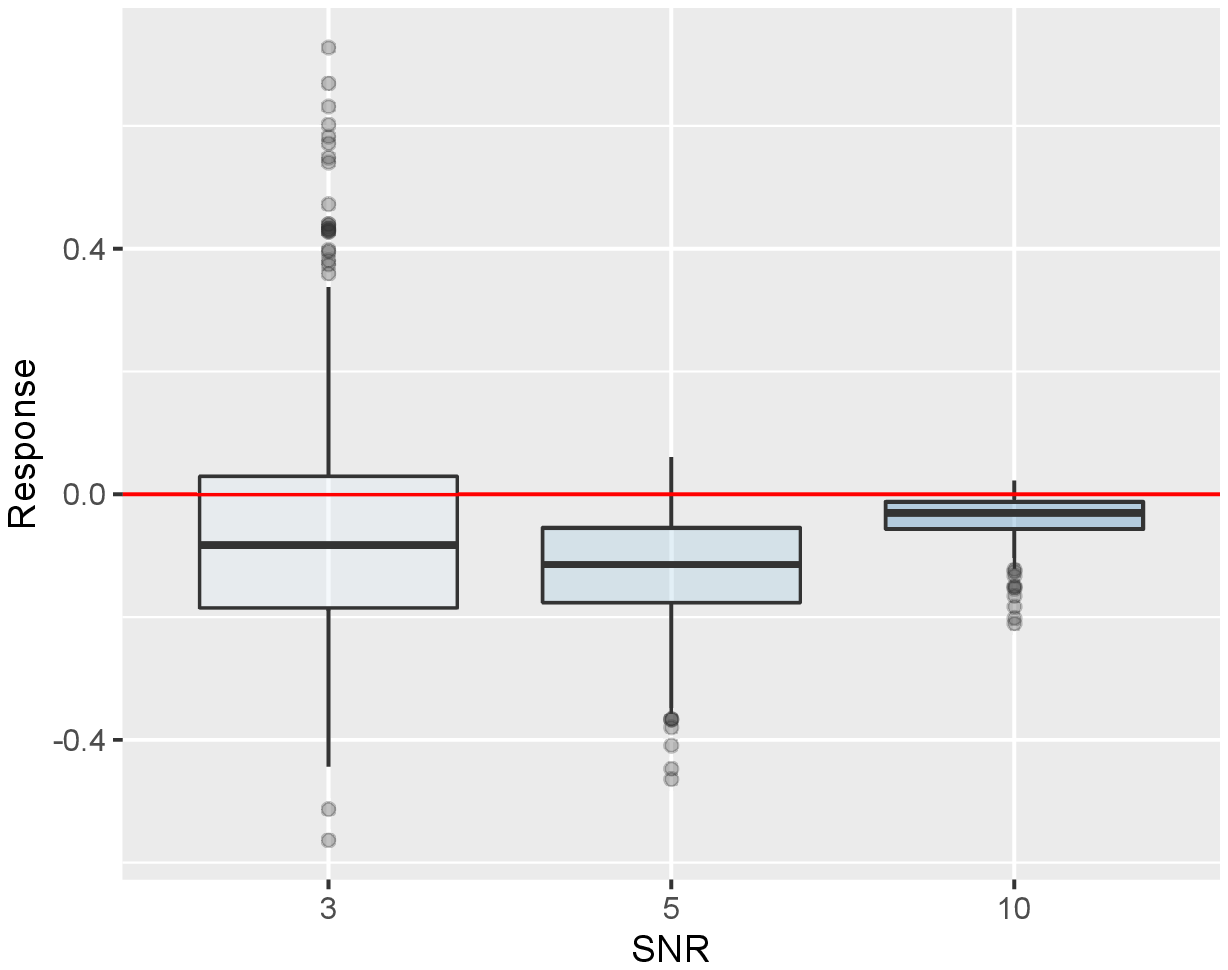}
		\caption{}
	\end{subfigure}%
	\begin{subfigure}{.50\textwidth}
		\centering
		\includegraphics[width=1\linewidth]{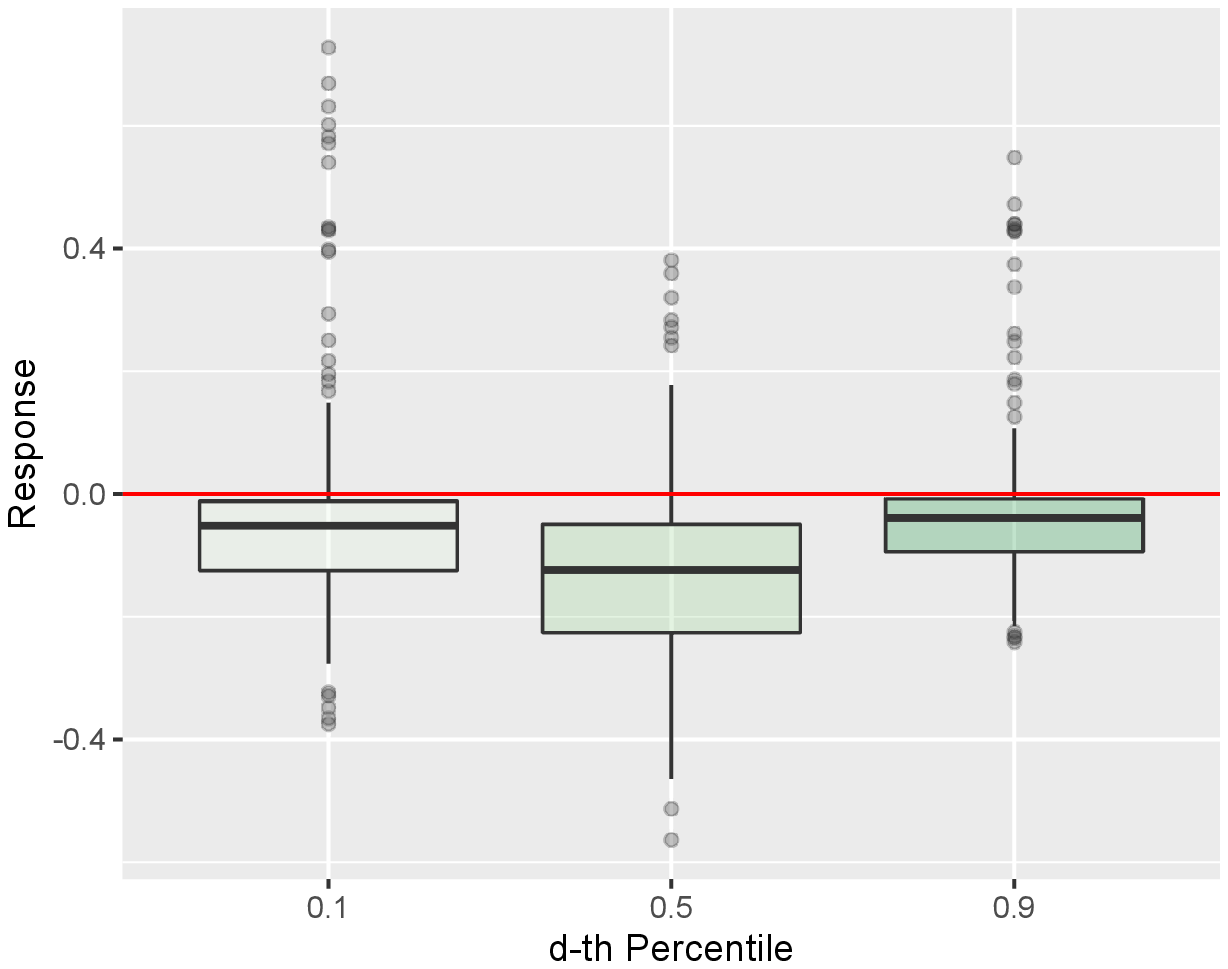}
		\caption{}
	\end{subfigure}
	\caption{Main effect plots of $SNR$ (a) and $d$-th Percentile (b).  The horizontal red solid line in each panel stands for Response equal to zero. Response that are negative indicates the truncated model outperforms BMA when it comes to squared errors.
	}
	\label{plot:main_effect}
\end{figure}

\begin{figure}[H]
	\centering
	\includegraphics[width=0.8\linewidth]{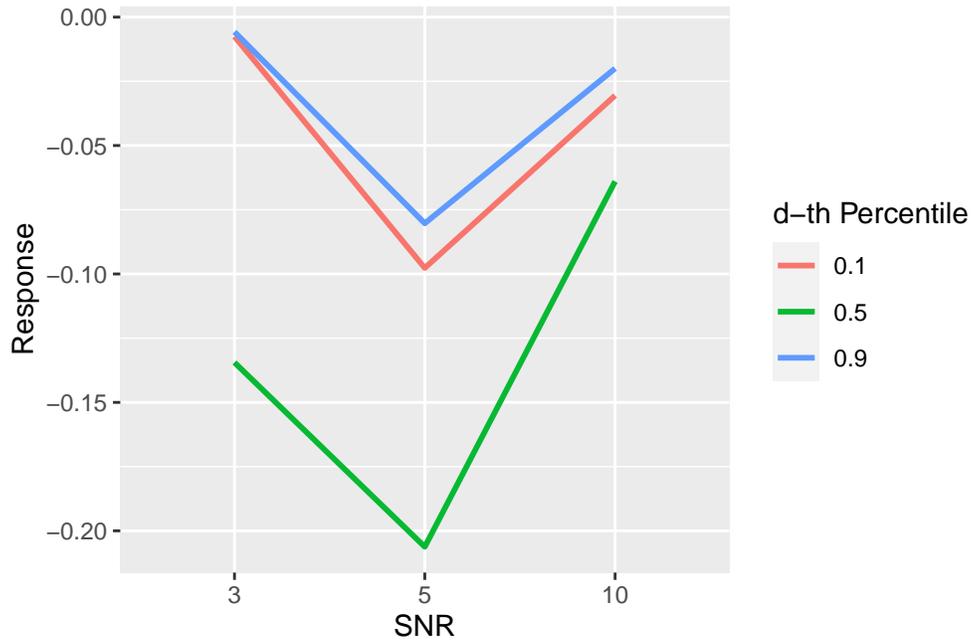}
	\caption{Plot of average of the Response by $SNR$ and $d$. Response that are negative implies the truncated model outperforms BMA regarding squared errors. The blue solid line indicates $d=0.9$, the green solid line indicates $d=0.5$, and the orange solid line indicates $d=0.1$.}
	\label{plot:interaction}
\end{figure}




\section{Real Data Analysis}


The American Community Survey (ACS) is an ongoing survey conducted by the U.S. Census Bureau annually and published on the website (\url{https://www.census.gov/programs-surveys/acs}). The purpose of ACS is to provide up-to-date estimates that are related to society and economy for a variety of geographies to the U.S. public. The U.S. Census Bureau launched the ACS in 2005. Since then, the public-use ACS estimates are released yearly on the basis of 1-year, 3-year, or 5-year periods. However, 3-year estimates, which were available for the areas with population greater than 20,000, were terminated in 2013. The 1-year estimates are accessible for the areas with at least 65,000 people, while no population restriction is put for the 5-year estimates. 

Motivated by an application of the ACS data, \cite{bradley2015spatio} proposed the Spatio-Temporal Change of Support (STCOS) methodology. This novel methodology was developed based on the fact that one may interested in getting estimates on spatial and/or temporal domains, which differ from the observed domains. The model results in a mixed effects model, where the coefficients of the random effects are structured to account for the multiple space/time scales. This in an example where Gaussian mixed effects model is used to analyze the data but there is no completing method in the literature (i.e. $B=1$). Thus, this application provides a good example of how our methodology can be used to obtain gains in prediction even though $B=1$. To illustrate our approach, we adopt the STCOS analysis of income data from \cite{unknown}. This dataset consists of all released 1-year, 3-year, and 5-year period ACS estimates of median household income over various geographies, such as conuties and census block-group level, within Missouri. The ACS estimates are consist of point estimates, margins of errors (MOE), and variance estimates. For this application, we adapt our methodology to the STCOS model applied to ACS median household income data recorded over the 2017 5-year period at the block-level to predict median household income at four neighborhoods in the Boone County, MO.


The truncated STCOS as a Bayesian hierarchical model can be written as 

\begin{align}
\label{real:model}
\nonumber
&\text{Data Model:	}\quad 
Z_i\mid Y_i,\bm{\theta},\sigma^2\sim \mathrm{N} \left(Y_i,\sigma^2\right) 
I\{\sum_{i}CPE_i[Z_i,\hat{Y}_i(\bm{\theta})]<\kappa\}
\\
\nonumber
&\text{Process Models:	}\quad 
Y_i\mid \bm{\beta},\bm{\eta},\sigma_{\xi}^{2}\sim \mathrm{N}\left(\bm{X}_i'\bm{\beta}+\bm{\psi}_i'\bm{\eta},
\sigma_{\xi}^{2} \right),\quad \bm{\eta}\mid\sigma_{K}^{2} \sim \mathrm{N}\left(\mathbf{0}, \sigma_{K}^{2} \textbf{K}\right)\\
\nonumber
&\text{Parameter Model 1:	}\quad 	\bm{\beta} \mid \sigma_{\mu}^{2} \sim \mathrm{N}\left(\bm{0}, \sigma_{\mu}^{2} \bm{I}\right)\\
\nonumber
&\text{Parameter Model 2:	} \quad \sigma_{\mu}^{2} \sim \mathrm{IG}\left(a_{\mu}, b_{\mu}\right)		
\\
\nonumber										
&\text{Parameter Model 3:	}
\quad \sigma_{K}^{2} \sim \mathrm{IG}\left(a_{K}, b_{K}\right) 
\\
&\text{Parameter Model 4:	}
\quad \sigma_{\xi}^{2} \sim \mathrm{IG}\left(a_{\xi}, b_{\xi}\right),
\end{align}
\noindent
where $\bm{X}_i = \left( \frac{\mid A_i \bigcap B_1 \mid}{\mid  A_i\mid},\dots, \frac{\mid  A_i \bigcap B_{n_B} \mid}{\mid  A_i\mid} \right)'$,  $\bm{\psi}_i=\left( \psi_1(A_i,{\ell}_i,t_i),\dots,\psi_r(A_i,{\ell}_i,t_i) \right)'$, $B_1,\dots, B_{n_B}$ are fine-scale grid points over the spatial domain, $\bm{\theta}=(\bm{\beta},\sigma_{\xi}^{2},\sigma_{K}^{2})'$, $\mid A\mid$ is denoted as the total surface area for areal unit $A$, $A_i$ is the areal unit associate with the $i$-th observation, ${\ell}_i$ is the period associated the $i$-th observation, $t_i$ is the time point associate with the $i$-th observation, $I(\cdot)$ is the indicator function, the matrix $\textbf{K}$ is a structure covariance matrix based on a random walk (details about this structure can be found in the paper of \cite{unknown}), and is multiplied with a free parameter $\sigma^2_K$ to fully define the covariance of the random coefficient $\bm{\eta}$. We have dropped $b$ in our notion for $\hat{Y}_i(\bm{\theta},b)$ because $B=1$. Set $\bm{\psi}_j(A,\ell,t)=\frac{1}{\ell\mid A \mid}\displaystyle\sum_{j=t-\ell+1}^{t}\underset{A}{\int}g_j(\bm{s},j)d\bm{s}$, where $g_j(\bm{s},j)$ represents a collection of spatio-temporal bisquare basis functions.

The STCOS model is a highly structured Bayesian mixed effect model for Gaussian data, where the random effect coefficients deliberate different spatio-temporal scales, and the covariates are the percentage of overlapping regions between the data's spatial support and a fine-scale grid. The purpose of this application is to show that our methodology can benefit prediction accuracy even when $B=1$.

\begin{figure}[H]
	\centering
	\includegraphics[width=0.8\linewidth]{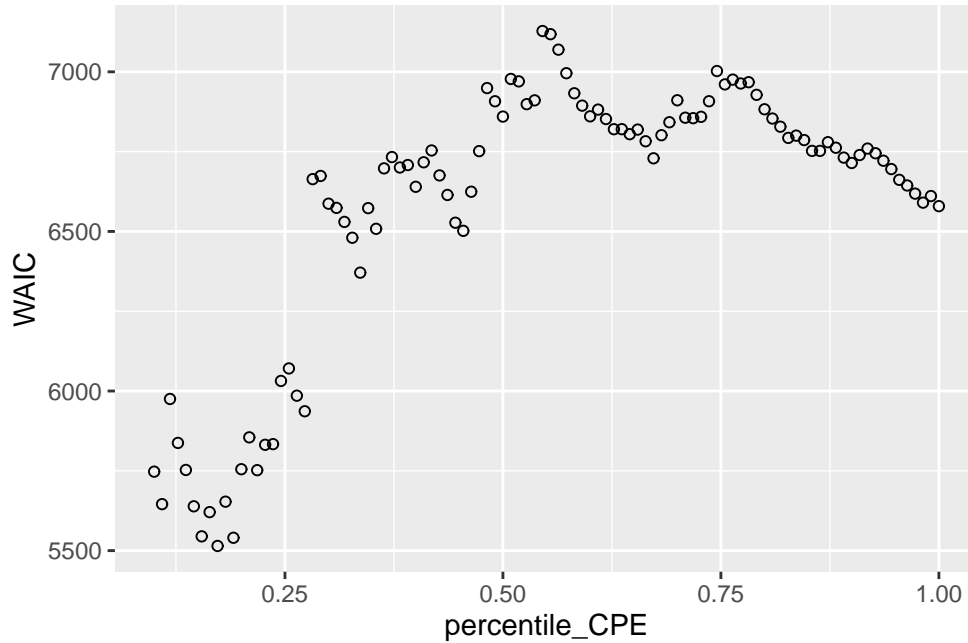}
	\caption{ $WAIC$ from the truncated model with $\kappa$ set equal to the $d$-th versus the percentiles of CPE from the untruncated model. Smaller value of $WAIC$ suggest better out-of-sample predictive accuracy.}
	\label{plot:waic}
\end{figure} 
To assess out-of-sample performance, we use the Wantanbe Akiake information criteria ($WAIC$). In Figure \ref{plot:waic},  $WAIC$ of the truncated model versus $\kappa$ set equal to different $d$-th percentile of $CPE$ from the model in (\ref{real:model}) wihtout any truncation. The sequence of $d$ is chosen to be $0.1$ to $1$ with a length of 100. $WAIC$ decreases as $d$ increases, then increases as $d$ increases, and is fairly constant between $0.5$ to $1$. When $\kappa$ is chosen to be the $0.17$ percentile, $WAIC$ reaches the smallest value. Therefore, we use $\kappa$ set to be the $0.17$ percentile of  CPE from (\ref{real:model}) wihtout any truncation for inference. When comparing  Table \ref{tbF} with Table \ref{tbR}, we see that predictions are fairly similar, but the measures of variability, in general, are larger for the untruncated model. Thus, this comparison, along with the $WAIC$ values in Figure \ref{plot:waic}, suggests that we may be outperforming the ``untruncated CPE model.'' 

\begin{table}[h]
	\centering
	\caption{Untruncated model-based estimates of 2017 median household income in four neighbors of Boone County: Central, East, North, and Paris}
	\begin{tabular}{|l|l|l|l|}
		\hline
		\multicolumn{1}{|l|}{Region} & Posterior Mean  & Posterior Standard Deviation    \\ \hline
		Central                      & 27047.33 & 1895.125 \\ \hline
		East                         & 43765.68 & 2453.249 \\ \hline
		North                        & 43483.82 & 2854.626 \\ \hline
		Paris63Corridor              & 19563.84 & 3910.908 \\ \hline
	\end{tabular}
\label{tbF}
\end{table}

\begin{table}[h]
	\centering
	\caption{Truncated model-based estimates of 2017 median household income in four neighbors of Boone County: Central, East, North, and Paris}
	\begin{tabular}{|l|l|l|l|}
		\hline
	\multicolumn{1}{|l|}{Region} &  Posterior Mean  & Posterior Standard Deviation   \\ \hline
	Central                      & 27005.93 & 1719.283 \\ \hline
	East                         & 43688.09 & 2442.975 \\ \hline
	North                        & 43243.18 & 2753.046 \\ \hline
	Paris63Corridor              & 19686.20 & 3941.126 \\ \hline
	\end{tabular}
\label{tbR}
\end{table}

\section{Discussion}
We propose a new approach towards model selection when either a single model or several candidate models are available for statistical inference. Instead of choosing a single model or combining the candidate models as is done in traditional model selection, we select a subset of the combined parameter space of the candidate model using an extension of an approach proposed by \cite{yekutieli2012adjusted}. Our new approach uses \cite{efron2004estimation} covariance penalized error (CPE) as a model selection criterion, and selects a subset of parameter space based on the values of CPE. Explicitly, the subset is formed by truncating the joint support of the data and the parameter space to only include small values of CPE. We show that Theorem \ref{Them:1} shows that our choice of truncation can lead to improvements of the mean squared prediction errors (MSPE) of predictors. We provide additional motivation for this truncated CPE model by showing that every Bayesian model for normal data can be interpreted as a type of truncated CPE model in Theorem \ref{Them:3}.

The simulation study shows that when we truncate half the MCMC replicates, after a burin-in, we obtain consistently better mean squared prediction errors than the original Bayesian model over  three different signal-to-noise specifications. The results also show that if you truncate too much or too little, we see little to no improvement on the basis of the squared errors. Hence, the selection of $\kappa$ appears to be an important choice, and in practice, we suggest using $WAIC$. The real data study of ACS period estimates demonstrates that prediction accuracy improvements can be achieved when applying our methodology to a single model. There are clear improvements to the out-of-sample error according to the $WAIC$, the values of the prediction in our study do not change substantially, but there are noticeable changes to the estimate of the variability.

The term $\kappa$ is an unknown parameter, and its specification can lead to either improvements or no changes. Thus, a natural extension of our method is to place a prior distribution on $\kappa$, as our use of the $WAIC$ to estimate $\kappa$ has unchecked variability not accounted for in the model. The theoretical results in this article may provide some guidance. For example, $\kappa^{*}$ in Theorem 2 follows a chi-square distribution, and the original Bayesian model is a re-scaled (to the power $r$) truncated CPE model with a chi-square prior placed on the upper bound. Thus, priors on $\kappa$ that imply a stochastic ordering relative to a chi-square distribution is an interesting topic of future research.

\section*{Acknowledgments} 
Jonathan R. Bradley's research was partially supported by the U.S. National Science Foundation (NSF) under
NSF grant SES-1853099 and the National Institute of Health (NIH) under grant 1R03AG070669-01.


\section*{Appendix A: Technical Results}
\subsection*{Proof of Thoerem \ref{Them:3}}
By definition
\begin{align*}
&	\pi(\bdm{y},\bm{\theta},b\vert \bdm{z}) = \frac{1}{\pi(\bdm{z})}f(\bdm{z}\mid \bdm{y},\sigma^2)\pi(\bdm{y}\vert \bm{\theta},b) \pi(\bm{\theta})\pi(b)h(\bdm{z},\bm{\theta},b),\\
\end{align*}
and when writing $f(\bdm{z}\mid \bdm{y},\sigma^2)=f(\bdm{z}\mid \bdm{y},\sigma^2)^rf(\bdm{z}\mid \bdm{y},\sigma^2)^{1-r}$,
\begin{align*}
\pi(\bdm{y},\bm{\theta},b\vert \bdm{z})=\frac{1}{\pi(\bdm{z})}f(\bdm{z}\mid \bdm{y},\sigma^2)^r f(\bdm{z}\mid \bdm{y},\sigma^2)^{1-r} \pi(\bdm{y}\vert \bm{\theta},b) \pi(\bm{\theta})\pi(b)h(\bdm{z},\bm{\theta},b),
\end{align*}
and introducing $u$ in a similar manner to \citet{damlen1999gibbs} such that,
\begin{align*}
\pi(\bdm{y},\bm{\theta},b\vert \bdm{z})=\frac{1}{\pi(\bdm{z})} \int_{0}^{1}
f(\bdm{z}\mid \bdm{y},\sigma^2)^{r}\pi(\bdm{y}\vert \bm{\theta},b) \pi(\bm{\theta})\pi(b)
I\{u<f(\bdm{z}\mid \bdm{y},\sigma^2)^{1-r}\}h(\bdm{z},\bm{\theta},b)du.
\end{align*}
Within the expression of the indicator take the log and multiply by -2 to obtain.
\begin{align*}
\pi(\bdm{y},\bm{\theta},b\vert \bdm{z})=\frac{1}{\pi(\bdm{z})} \int
f(\bdm{z}\mid \bdm{y},\sigma^2)^{r}\pi(\bdm{y}\vert \bm{\theta},b) \pi(\bm{\theta})\pi(b)
I\{-2(1-r)log(f(\bdm{z}\mid \bdm{y},\sigma^2))<-2log(u)\}h(\bdm{z},\bm{\theta},b)du.
\end{align*}
Then, upon substituting the expression $r =\frac{CPE(\bdm{z},\hat{\bdm{y}})}{2log\left(f(\bdm{z}\vert \bdm{y},\sigma^{2})\right)} + 1$, we obtain the result.

\subsection*{Proof of Thoerem \ref{Them:1}}
By construction $CPE(\bdm{z},\hat{\bdm{y}}_{tc})<\kappa$. For $\kappa=E\{\sum_{i=1}^{n}(Y_i-\hat{Y}_{i,m})^2\}+n\sigma ^2$, we have 
\begin{equation}
\label{appen:a}
CPE(\bdm{z},\hat{\bdm{y}}_{tc})<E\{\sum_{i=1}^{n}(Y_i-\hat{Y}_{i,m})^2\}+n\sigma ^2.
\end{equation}
By stein's lemma (\citep{stein1981estimation}), upon taking the expected value across Expression (\ref{appen:a}), we obtain the result.
\subsection*{Proof of Corollary \ref{Them:2}}
It follows from Theorem 2 that,
\begin{equation*}
\frac{1}{n}\sum_{i = 1}^{n}E\{Y(\bdm{s}_{i})-\hat{Y}_{tc}(\bdm{s}_{i})\}^2< \frac{1}{n}\sum_{i = 1}^{n}E\{Y(\bdm{s}_{i})-\hat{Y}_{m}(\bdm{s}_{i})\}^2.
\end{equation*}
Then apply the law large numbers \citep{billingsley2013convergence} as $n$ approaches infinity to obtain the result.
\section*{Appendix B: Derivation of full-conditional distributions for Gibbs Sampling}
Let $\bdm{C}(b,\tau^2)=\tau^2\bdm{H}(b)$, and let $ \bdm{y}= \bdm{X}\bm{\beta}+\bdm{w}$, where $\bdm{w}|\tau^2,b\sim N(\bm{0},\bdm{C}(b,\tau^2))$. In our Gibbs sampler, we update $\bm{\beta}$, and $\bdm{w}$, which implicitly updates $\bdm{y}$. We provide the derivations of the full-conditional distributions associated with the model in (\ref{simu:model}) with a bulleted list as follows.\\
$\bullet$ Full-conditional distribution for $\bdm{w}$:
\begin{align*}
f(\bdm{w}|\cdot) &\propto 
exp\left\lbrace-\frac{(\bdm{z}-\bdm{X}\bm{\beta}-\bdm{w})'(\bdm{z}-\bdm{X}\bm{\beta}-\bdm{w})}{2\sigma^2}-\frac{\bdm{w}'\bdm{H}(b)^{-1}\bdm{w}}{2\tau^2}\right\rbrace I\{CPE[\bdm{z},\hat{\bdm{y}}(\bm{\theta},b)]<\kappa\}\\
&\propto exp\left\lbrace\frac{-\bdm{w}'\bdm{w}}{2\sigma^2}+\frac{2\bdm{w}'(\bdm{z}-\bdm{X}\bm{\beta})}{2\sigma^2}-\frac{\bdm{w}'\bdm{H}(b)^{-1}\bdm{w}}{2\tau^2}\right\rbrace I\{CPE[\bdm{z},\hat{\bdm{y}}(\bm{\theta},b)]<\kappa\}\\
&\propto
exp\left[\frac{-\bdm{w}'\left\lbrace\frac{1}{\sigma^2}\bdm{I}_n+\frac{1}{\tau^2}\bdm{H}(b)^{-1}\right\rbrace\bdm{w}}{2}+\frac{2\bdm{w}'(\bdm{z}-\bdm{X}\bm{\beta})}{2\sigma^2}\right] I\{CPE[\bdm{z},\hat{\bdm{y}}(\bm{\theta},b)]<\kappa\}\\
&\propto
exp\left\lbrace\frac{-\bdm{w}'\bm{\Sigma}_w^{-1}\bdm{w}}{2}+	\frac{2\bdm{w}'\bm{\Sigma}_w^{-1}\bm{\Sigma}_w(\bdm{z}-\bdm{X}\bm{\beta})}{2\sigma^2}\right\rbrace I\{CPE[\bdm{z},\hat{\bdm{y}}(\bm{\theta},b)]<\kappa\}\\
&\propto
exp\left(\frac{-\bdm{w}'\bm{\Sigma}_w^{-1}\bdm{w}}{2}+	\frac{2\bdm{w}'\bm{\Sigma}_w^{-1}\bm{\mu}_w}{2}\right)I\{CPE[\bdm{z},\hat{\bdm{y}}(\bm{\theta},b)]<\kappa\}\\
&\propto
N(\bm{\mu}_w,\bm{\Sigma}_w) I\{CPE[\bdm{z},\hat{\bdm{y}}(\bm{\theta},b)]<\kappa\},
\end{align*}
where $\bm{\mu}_w=\frac{\bm{\Sigma}_w(\bdm{z}-\bdm{X}\bm{\beta})}{\sigma^2}, \bm{\Sigma}_w=\frac{1}{\sigma^2}\bdm{I}_n+\frac{1}{\tau^2}\bdm{H}(b)^{-1} $.\\
$\bullet$ Full-conditional distribution for $\bm{\beta}$:
\begin{align*}
f(\bdm{\beta}|\cdot) &\propto
exp\left\lbrace\frac{-(\bdm{z}-\bdm{X}\bm{\beta}-\bdm{w})'(\bdm{z}-\bdm{X}\bm{\beta}-\bdm{w})}{2\sigma^2} -\frac{\bm{\beta}'\bm{\beta}}{20}\right\rbrace I\{CPE[\bdm{z},\hat{\bdm{y}}(\bm{\theta},b)]<\kappa\}\\
&\propto exp\left\lbrace\frac{-\bm{\beta}'\bdm{X}'\bdm{X}\bm{\beta}}{2\sigma^2}+
\frac{2\bm{\beta}'\bdm{X}'(\bdm{z}-\bdm{w})}{2\sigma^2}-\frac{\bm{\beta}'\bm{\beta}}{20}\right\rbrace I\{CPE[\bdm{z},\hat{\bdm{y}}(\bm{\theta},b)]<\kappa\}\\&\propto
exp\left\lbrace\frac{-\bm{\beta}'(\frac{\bdm{X}'\bdm{X}}{\sigma^2}+\frac{1}{10}\bdm{I}_p)\bm{\beta}}{2}+\frac{2\bm{\beta}'\bdm{X}'(\bdm{z}-\bdm{w})}{2\sigma^2}
\right\rbrace I\{CPE[\bdm{z},\hat{\bdm{y}}(\bm{\theta},b)]<\kappa\}\\
&\propto
exp\left\lbrace\frac{-\bm{\beta}'\bm{\Sigma}_{\beta}^{-1}\bm{\beta}}{2}+\frac{2\bm{\beta}'\bm{\Sigma}_{\beta}^{-1}\bm{\Sigma}_{\beta}\bdm{X}'(\bdm{z}-\bdm{w})}{2\sigma^2}\right\rbrace I\{CPE[\bdm{z},\hat{\bdm{y}}(\bm{\theta},b)]<\kappa\}\\
&\propto
N(\bm{\mu}_{\beta},\bm{\Sigma}_{\beta}) I\{CPE[\bdm{z},\hat{\bdm{y}}(\bm{\theta},b)]<\kappa\},
\end{align*}
where $\bm{\mu}_{\beta}=\frac{\bm{\Sigma}_{\beta}\bdm{X}'(\bdm{z}-\bdm{w})}{\sigma^2},\bm{\Sigma}_{\beta}=\frac{\bdm{X}'\bdm{X}}{\sigma^2}+\frac{1}{10}\bdm{I}_p $.\\
$\bullet$ Full-conditional distribution for $\tau^2$:
\begin{align*}
f(\tau^2|\cdot) &\propto
\frac{1}{|\tau^2\bdm{H}(b)|^{1/2}} exp\left[\frac{-\bdm{w}'\left\lbrace\frac{1}{\tau^2}\bdm{H}(b)^{-1}\right\rbrace\bdm{w}}{2}-\frac{0.01}{\tau^2}\right]\left(\frac{1}{\tau^2}\right)^{1/2}\\
&\propto  exp\left[\frac{-\frac{1}{2}\bdm{w}'\bdm{H}(b)^{-1}\bdm{w}-0.01}{\tau^2}\right]\left(\frac{1}{\tau^2}\right)^{n/2+2}\\	
&\propto
IG(0.5n+1,0.01+0.5\bdm{w}'\bdm{H}(b)^{-1}\bdm{w}),
\end{align*}
where $|\bdm{H}(b)|$ is the determinant of $\bdm{H}(b)$.\\
$\bullet$ Full-conditional distribution for $b$:		
\begin{align*}
f(b=j|\cdot)&=\frac{ exp\left[\frac{-\bdm{w}'\bdm{H}(b=j)^{-1}\bdm{w}}{2\tau^2}\right]|\bdm{H}(b=j)|^{-1/2}}
{\sum_{k=1}^{6}exp\left[\frac{-\bdm{w}'\bdm{H}(b=k)^{-1}\bdm{w}}{2\tau^2}\right]|\bdm{H}(b=k)|^{-1/2}
}
\end{align*}

\bibliographystyle{jasa}  
\bibliography{myref_1}

 \end{document}